\documentclass[aps,prl,twocolumn,groupedaddress,floatfix]{revtex4}

\usepackage{graphicx}% Include figure files
\usepackage{subfigure}% Include subfigures
\usepackage{epsfig}
\usepackage{amsmath}
\usepackage{bbm}

\def\be{\begin{equation}}
\def\ee{\end{equation}}
\def\bea{\begin{eqnarray}}
\def\eea{\end{eqnarray}}

\begin{document}

% Use the \preprint command to place your local institutional report
% number in the upper righthand corner of the title page in preprint mode.
% Multiple \preprint commands are allowed.
% Use the 'preprintnumbers' class option to override journal defaults
% to display numbers if necessary
%\preprint{}

%Title of paper
\title{A theory of superconductivity in multi-walled 
carbon nanotubes}

% repeat the \author .. \affiliation  etc. as needed
% \email, \thanks, \homepage, \altaffiliation all apply to the current
% author. Explanatory text should go in the []'s, actual e-mail
% address or url should go in the {}'s for \email and \homepage.
% Please use the appropriate macro foreach each type of information

% \affiliation command applies to all authors since the last
% \affiliation command. The \affiliation command should follow the
% other information
% \affiliation can be followed by \email, \homepage, \thanks as well.
\author{E. Perfetto$^{1,2}$ and J. Gonz{\'a}lez$^1$}
%\email[]{Your e-mail address}
%\homepage[]{Your web page}
%\thanks{}
%\altaffiliation{}
\affiliation{$^{1}$Instituto de Estructura de la Materia.
        Consejo Superior de Investigaciones Cient{\'\i}ficas.
        Serrano 123, 28006 Madrid. Spain.\\
             $^{2}$Istituto Nazionale di Fisica Nucleare - Laboratori
        Nazionali di Frascati, Via E. Fermi 40, 00044 Frascati, Italy.}
%Collaboration name if desired (requires use of superscriptaddress
%option in \documentclass). \noaffiliation is required (may also be
%used with the \author command).
%\collaboration can be followed by \email, \homepage, \thanks as well.
%\collaboration{}
%\noaffiliation

\date{\today}

\begin{abstract}
We devise an approach to describe the electronic instabilities
of doped multi-walled nanotubes, where each shell has in general a 
manifold of Fermi points. Our analysis relies on the scale dependence of
the different scattering processes, showing that a pairing 
instability arises for a large enough number of Fermi points 
as a consequence of their particular geometric arrangement. The instability 
is enhanced by the tunneling of Cooper pairs between nearest
shells, giving rise to a transition from the Luttinger liquid to a 
superconducting state in a wide region of the phase diagram.

\end{abstract}
% insert suggested PACS numbers in braces on next line
% insert suggested keywords - APS authors don't need to do this
%\keywords{}

%\maketitle must follow title, authors, abstract, \pacs, and \keywords
\maketitle
% body of paper here - Use proper section commands
% References should be done using the \cite, \ref, and \label commands
% Put \label in argument of \section for cross-referencing
%\section{\label{}}
%\subsection{}
%\subsubsection{}
%%%%%%%%%%%%%%%%%%%%%%%%%%%%%%%%%%%%%%%%%%%%%%%%%%%%%%%%%%%%%%%%%%%
%%%%%%%%%%%%%%%%%%%%%%%%%%%%%%%%%%%%%%%%%%%%%%%%%%%%%%%%%%%%%%%%%%%

Carbon nanotubes offer nowadays a great potential for the investigation
of mesoscopic physics. Their reduced dimensionality leads to very strong 
electron correlations, with the consequent breakdown of the conventional
Fermi liquid picture\cite{bal,eg,kane,yo}. 
This has become manifest, for instance, in the 
experimental observation of a power-law suppression of the conductance 
over a wide range of temperatures\cite{exp,yao}. 
In a different type of experiments, superconducting (SC) correlations have 
been also observed in carbon nanotubes at low temperatures. 
There have been measures of a sharp decrease in the resistance of 
ropes with a large number ($\sim 300$) of nanotubes, with critical 
temperatures $T_c \sim 1$ K \cite{kas,sup}. 
More recently, quite abrupt drops in the resistance have been measured in 
large arrays of multi-walled nanotubes (MWNTs) grown in the pores of an 
alumina template\cite{takesue}, with transition temperatures 
$T_c \sim 10 \; {\rm K}$.

It has been shown that the growth of the SC correlations becomes possible
in the carbon nanotubes when the screened Coulomb interaction is overcome 
by the effective interaction mediated by phonons\cite{th1,th2}. For this 
to happen, it is required in general the coupling between a large number 
of conducting channels, which favors the screening of the Coulomb 
interaction. The case of the MWNTs is special in that a certain amount 
of doping seems to be required, to ensure the metallic character of all 
the shells and the possibility of Cooper-pair tunneling between them.
This is consistent with the fact that the drops in the resistance have 
been only observed in samples where all the shells in the MWNTs are contacted
by the electrodes\cite{takesue}. 
%In this perspective, the superconductivity of the MWNTs may be more related
%to that of the doped compounds of graphite, which have in some cases 
%critical temperatures of the order of $\sim 10$ K \cite{nemery}.

\begin{figure}
\begin{center}
\mbox{\epsfxsize 6.0cm \epsfbox{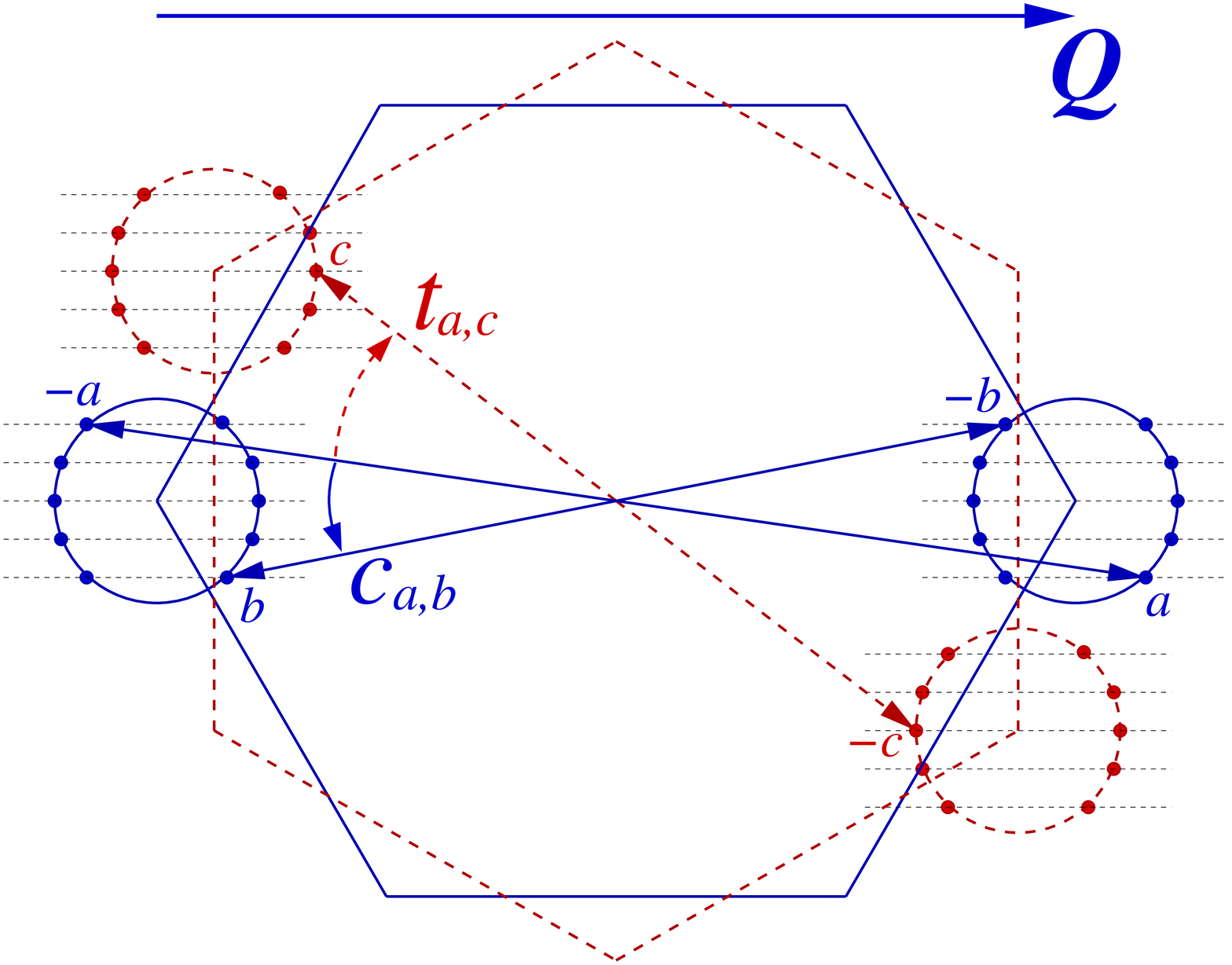}}
\end{center}
\caption{Schematic representation of the location of the Fermi points for the
shells of a doped MWNT, referred to the hexagonal Brillouin zone (BZ) of graphene.
The horizontal lines correspond to different low-energy subbands, arising from the
conical dispersion around the corners of the BZ. The full (dashed) hexagon stands for
the case of an armchair (zig-zag) geometry. The $c_{a,b}$ ($t_{a,b}$) couplings
label the intratube (intertube) scattering of Cooper pairs at zero total momentum.}
\label{fermipts}
\end{figure}

The aim of this paper is to study the low-energy instabilities of doped 
MWNTs, which may have a large number of subbands crossing the Fermi level 
in each shell. These give rise to a manifold of Fermi points located in two 
disconnected patches around the $K$ and $K'$ points in momentum space, 
with an approximate circular shape as represented in Fig. \ref{fermipts}.
For typical shells with radius $R \sim 10 \; {\rm nm}$,
the energy spacing between the subbands is $\Delta E \sim 0.1 \; {\rm eV}$.
Below this energy, the different excitations are given by approximate
linear branches around each Fermi point, corresponding to the dispersion in
the longitudinal momentum $p$. The electronic properties are dictated then by 
the interaction among a collection of one-dimensional (1D) electron liquids, 
with electron fields $\psi_a (p)$ classified by a Fermi point index $a$ that,
in our notation, changes sign upon inversion in momentum space. 
In this regime, the scattering processes are severely restricted 
by momentum conservation, and the relevant interactions can be read from the 
hamiltonian
\begin{eqnarray}
H_{\rm int}   &  =  &  \int dp dp' dq  
  \left[ \psi^{+}_{a} (p+q) \psi_{a} (p) \; f^{(+)}_{a,-b} \;
             \psi^{+}_{-b} (p'-q) \psi_{-b} (p')  
                                                \right.    \nonumber    \\
  &  &  +  \psi^{+}_{-b} (p+q) \psi_{a} (p)  \; f^{(-)}_{a,-b} \; 
             \psi^{+}_{a} (p'-q) \psi_{-b} (p')      \nonumber    \\
  &  &  +  \psi^{+}_{b} (p+q) \psi_{a} (p)  \; c^{(+)}_{a,b} \;
             \psi^{+}_{-b} (p'-q) \psi_{-a} (p')      \nonumber    \\
  &  &   \left.   +  \psi^{+}_{-b} (p+q) \psi_{a} (p)  \; c^{(-)}_{a,b} \;
             \psi^{+}_{b} (p'-q) \psi_{-a} (p')    \right]  
\label{hint}
\end{eqnarray}
where there is an implicit sum over all the pairs of Fermi points $a, b$ with 
like chirality (i. e. right- or left-moving character).
We observe that, as in 2D systems with a continuous Fermi line, all the
relevant interactions can be classified into forward-scattering ($f^{(+)}_{a,b}$),
exchange ($f^{(-)}_{a,b}$), and Cooper-pair ($c^{(\pm )}_{a,b}$) channels\cite{lbf,sh}.

At the energy scale $\Delta E $, the main contribution to the different
couplings in (\ref{hint}) comes from the Coulomb potential, which in the 
nanotube geometry is given by
\begin{equation}
V_C ({\bf r}-{\bf r}')=\frac{ e^{2} /\kappa }
 {\sqrt{(x-x')^{2}+4R^{2}\sin^{2}[(y-y')/2R]+a_{z}^{2} }}
\label{potent}
\end{equation}
with $a_{z} \simeq 1.6 \; {\rm \AA}$  and $\kappa \approx 2.4$ \cite{eg}.
The Fourier transform of (\ref{potent}) dictates the value of the couplings 
$f^{(\pm )}_{a,b}$ and $c^{(\pm )}_{a,b}$, according to the corresponding
momentum-transfer. Thus, couplings with momentum-transfer of the order 
of the large momentum ${\bf Q}$ connecting the $K$ and $K'$ points
have an intrinsic $1/R$ dependence, inherent to the behavior of
the potential at large momentum. Moreover, in the case of the 
$f^{(-)}_{a,b}$ couplings, the different symmetry of the Bloch wave functions 
for left- and right-moving electrons leads also to an effective $1/R$ 
reduction of their strength\cite{bal,eg,kane}. The most important effects
of the Coulomb repulsion arise in the forward-scattering interactions, 
otherwise screened by particle-hole excitations about each of
the Fermi points in the different shells. This is consistent with the 
observation of Luttinger liquid behavior and the relative reduction of
the exponent for the power-law dependence of the conductance in doped 
MWNTs\cite{mw,egger}.

At energies much lower than $\Delta E$, however, the couplings grow
large in general, pointing at the onset of an electronic instability in 
the system. The character of this instability depends on whether there
are different shells electrically active in the MWNTs. 
Typical MWNTs are made of shells with a random mixture of helicities. 
In these conditions, the single-electron tunneling between nearest
shells is largely suppressed since, in general, the misalignment of
the respective carbon lattices leads to a relative rotation and to the
mismatch of the Fermi points\cite{mkm}, as shown in Fig. \ref{fermipts}.
However, such a mismatch does not prevent the intertube tunneling of 
Cooper pairs with vanishing total momentum. 
%This effect may become 
%quite important when all the shells participate in the conduction, 
%since it may enhance the superconducting correlations in the MWNTs. 
We will take this into account by introducing new intershell tunneling
couplings $t^{(\pm)}_{a,b}$, standing for the amplitude of a pair (with
electrons around Fermi points $a$ and $-a$) to tunnel into a neighbor 
shell (with electrons around $b$ and $-b$). To discern between
the possible tunneling of singlet or triplet pairs, we assign an 
amplitude $t^{(-)}_{a,b}$ to the process where the chiralities of the 
particles in the pair are exchanged, different to that for the direct 
process $t^{(+)}_{a,b}$.

Below the energy scale $\Delta E$, the electron system is effectively
1D, and the interaction vertices are corrected by diagrams
that depend logarithmically on the energy scale $\Lambda $ measured 
from the Fermi energy $\varepsilon_F $. To deal with this divergence, 
one can integrate progressively the electron modes starting from the 
limits of the linear branches at $\varepsilon_F \pm \Delta E $ \cite{sol}. 
This leads to a dependence of the couplings on the scale 
$l = -\log (\Lambda / \Delta E)$, according to the scaling equations
\begin{eqnarray}
\frac{\partial f^{(+)}_{a,b} }{ \partial l}  &  =  &  
  -\frac{1}{2\pi v_{ab}} (  (f^{(-)}_{a,b})^2  -  (c^{(+)}_{a,-b})^2  )     \\
\label{fs1}
\frac{\partial f^{(-)}_{a,b} }{ \partial l}  &  =  &
   -\frac{1}{\pi v_{ab}} (  (f^{(-)}_{a,b})^2  +  (c^{(-)}_{a,-b})^2  
             -  c^{(-)}_{a,-b}  c^{(+)}_{a,-b}  )
\end{eqnarray}
where the $v_{ab}$ are defined in terms of the 
string of Fermi velocities $v_a$ as $v_{ab} = (v_a + v_b)/2$. 

The couplings $c^{(\pm)}_{a,b}$ are also renormalized by the intertube
tunneling amplitudes $t^{(\pm)}_{a,b}$, and their scaling equations turn
out to be
\begin{eqnarray}
\frac{\partial c^{(+)}_{a,b} }{ \partial l}  &  =  &  
   -\sum_{c,s}   \frac{1}{2\pi v_c}
      ( c^{(s)}_{a,c} c^{(s)}_{c,b} + t^{(s)}_{a,c} t^{(s)}_{c,b} )
    +  \frac{1}{2\pi v_{ab}}   c^{(+)}_{a,b} h^{(+)}_{a,-b} 
\label{coop1}      \\
\frac{\partial c^{(-)}_{a,b} }{ \partial l}  &  =  &
   -\sum_{c,s}    \frac{1}{2\pi v_c}
     ( c^{(s)}_{a,c} c^{(-s)}_{c,b} + t^{(s)}_{a,c} t^{(-s)}_{c,b} ) 
                                                        \nonumber         \\
  &  &     -  \frac{1}{\pi v_{ab}}     c^{(-)}_{a,b} h^{(-)}_{a,-b}
   -   \frac{1}{2\pi v_{ab}}     \sum_{s}  c^{(s)}_{a,b}  h^{(-s)}_{a,-b} 
\label{coop2}            \\
\frac{\partial t^{(\pm)}_{a,b} }{ \partial l}  &  =  &
     -\sum_{c,s}    \frac{1}{2\pi v_c}
    ( c^{(s)}_{a,c} t^{(\pm s)}_{c,b}  +  t^{(s)}_{a,c} c^{(\pm s)}_{c,b} ) 
\label{coop3}
\end{eqnarray}
with $h^{(s)}_{a,-b} \equiv  2 f^{(s)}_{a,-b} -  \delta_{ab} c^{(s)}_{a,a}$.

The above equations apply to the description of MWNTs where all the shells
participate in the conduction, as the $t^{(\pm)}_{a,b}$ couplings imply
the transport between nearest shells. Their relative strength can be
estimated by taking the square of the interlayer single-electron hopping in
graphite ($\sim 0.1 \; {\rm eV}$), in units of the hopping rate in the graphene 
lattice ($\sim 2.5 \; {\rm eV}$), leading to relative amplitudes of the order 
of $\sim 0.002$. When the system has some tendency to develop superconducting 
correlations, such relatively small couplings at the scale $\Delta E $ are 
however reinforced at low energies, together with the Cooper-pair couplings.
An instability appears then in the electron system at scales which are 
typically in the range of $l \approx 4-6$.
This may be confronted with the instance where only one shell 
is electrically active, as it happens in most experiments with MWNTs. Then
the $t^{(\pm)}_{a,b}$ couplings have to be taken identically zero. The 
flow of Eqs. (\ref{fs1})-(\ref{coop3}) still has some unstable regime, but 
it appears at values of the variable $l \gtrsim 10$, which correspond to too
low temperatures to allow the observation of a new phase (given the 
appearance of Coulomb blockade effects in real samples at higher 
temperatures). 

We have seen that the contribution of the $e$-$e$ interaction to the 
intratube couplings at the scale $\Delta E$ is defined exclusively in terms 
of the potential (\ref{potent}), appropriately screened by the RPA sum of 
finite particle-hole processes preserving chirality about each of the Fermi 
points. To complete the evaluation of the couplings, we have 
added the contribution of the effective phonon-mediated interaction. 
We have relied on the results obtained for nanotubes in a wide range of 
diameters\cite{cara,conn}, which coincide in that the 
coupling $\lambda $ of the effective interaction for small momentum-transfer 
is about three times smaller (in absolute value) than that for momentum 
about ${\bf Q}$ \cite{graph}. We have thus taken a contribution 
to the original couplings from phonon-exchange that scales inversely 
proportional to the nanotube radius\cite{marti}, running between 
$\lambda = - 0.1 / R$ and $- 0.03 / R$ (with $R$ expressed in \AA  ) for 
the channels with small momentum-transfer, and respective values equal to 
$3 \lambda $ for the channels with momentum-transfer about 
${\bf Q}$\cite{cara,conn}. We have observed anyhow that the phase diagram 
of the system is quite insensitive to the variation of $\lambda $ within
that range. This is related to the fact that, as explained below, the 
dominant SC instability arises essentially from the anisotropy of the 
interactions, created by the peculiar geometry of disconnected patches of
Fermi points.

\begin{figure}
\begin{center}
\mbox{\epsfxsize 7.5cm \epsfbox{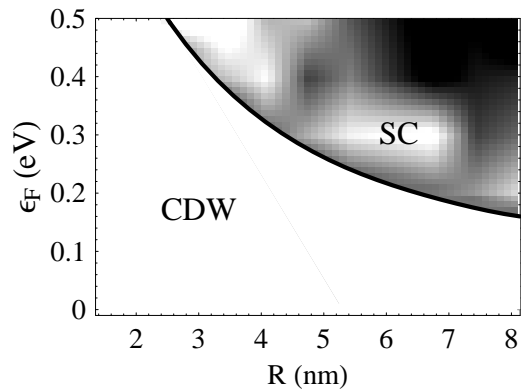}}
\end{center}
\caption{Phase diagram of doped MWNTs in terms of the average radius $R$ 
of the shells and the doping level, represented by the shift 
in the Fermi energy $\varepsilon_F$ with respect to the value in the 
undoped system. The upper region corresponds to 
a SC phase with $p$-wave symmetry. The density plot 
represents in grey scale the values of the logarithm of the transition scale 
$\omega_c $, ranging from $-\log (\omega_c / \Delta E ) \approx 4$ 
(black) to $\approx 6$ (white).}
\label{phases}
\end{figure}

We have characterized the low-energy electronic instabilities by looking
at the growth as $\Lambda \rightarrow 0$ of the different response functions, 
for operators corresponding to order parameters for all types of symmetries 
(charge-density-wave (CDW), spin-density-wave, superconductivity with various 
wave symmetries). The SC response functions are special in that, for a given 
$g$-wave symmetry, there is the intratube response function $R_g (\omega )$ and 
the intertube counterpart $T_g (\omega )$ measuring the propagation of Cooper
pairs between nearest shells. The respective derivatives with respect to the
frequency, $\overline{R}_g = \partial R_g / \partial \log \omega $ and 
$\overline{T}_g = \partial T_g / \partial \log \omega $, satisfy scaling 
equations in the usual fashion\cite{sol}
\begin{eqnarray}
\frac{\partial \overline{R}_g (\Lambda ) }{\partial \log (\Lambda )}
  &  =  &    \sum_{a,b,s} g^{(s)}_{a,b} \: c^{(s)}_{a,b} (\Lambda ) \:
                   \overline{R}_g (\Lambda )                              \\
\frac{\partial \overline{T}_g (\Lambda ) }{\partial \log (\Lambda )}
  &  =  &    \sum_{a,b,s} g^{(s)}_{a,b} \: t^{(s)}_{a,b} (\Lambda ) \:
                   \overline{R}_g (\Lambda )
\label{prop}
\end{eqnarray}
where $g^{(s)}_{a,b}$ is the symmetry factor appropriate for
each particular wave. We have observed that there is a region of the phase
diagram, represented in Fig. \ref{phases}, where the growth of the SC response 
functions is overshadowed by the divergence of any of the CDW response functions 
at low energies. Above the phase boundary shown in the figure, however, the 
CDW correlations are small and the instability is dictated by the large growth of 
both $R_g (\omega )$ and $T_g (\omega )$ at low energies. This is the signal that 
the Cooper pairs are able to propagate across the entire multi-walled structure. 
The singularity of $T_g (\omega )$ at some finite frequency $\omega_c $
marks the transition to a SC state, that we have always found in a channel of 
$p$-wave symmetry (together with subdominant $d$-wave pairing correlations).

Quite remarkably, the boundary between the CDW and the SC phase
in the diagram of Fig. \ref{phases} corresponds to the situation where the 
number of Fermi points changes from 12 to 20 in each shell of our model. 
We have solved the scaling equations for a number of up to 
$\sim 50$ Fermi points, which already demands the consideration of 
$\sim 1,000$ independent couplings. Then, we have found that the scale 
$\omega_c $ of the SC transition ranges between values corresponding to 
$-\log (\omega_c /\Delta E) \approx 4-6$. Recalling that
$\Delta E \sim 0.1 \; {\rm eV}$, we see that the predicted critical 
temperatures reach a higher scale of $\sim 10$ K, which is consistent
with the experimental observations\cite{takesue}.

There is a natural explanation for the appearance of SC correlations with 
$p$-wave symmetry, which relies on the peculiar arrangement of the Fermi 
points into two circular patches. These have a symmetry that corresponds to 
the group $C_{2v}$. The Cooper-pair vertex, taken as a 
function of the angular variables $\theta $ and $\theta ' $ of the 
pairs before and after the scattering, may be expanded in the form
\begin{eqnarray}
V(\theta, \theta ')   &  = &   V_0 + V_1 \cos (\theta) \cos (\theta ') 
 + V_2 \sin (\theta) \sin (\theta ')              \nonumber        \\
   &   &    + V_3 \cos (2\theta) \cos (2\theta ')
                     + V_4 \sin (2\theta) \sin (2\theta ')
\end{eqnarray}
A simple operation allows to estimate the different coefficients in the 
expansion. By placing one of the electrons in the incoming pair at the
maximum angle $\theta_0 $ spanned by one circular patch, we may parametrize
$V_1, V_2, V_3$ and $V_4$ in terms of the values of the vertex at zero 
momentum-transfer, $V(\theta_0, \theta_0) \equiv U_1$, at ${\bf Q}$ 
momentum-transfer, $V(\theta_0, \pi - \theta_0) \equiv U_2$, with incoming 
and outgoing electrons at opposite angles, 
$V(\theta_0, - \theta_0) \equiv U_3$, and at antipodal points, 
$V(\theta_0, \pi + \theta_0) \equiv U_4$. We obtain 
\begin{eqnarray}
V_1 \cos^2 (\theta_0)  &  =  &  (U_1 - U_2 + U_3 - U_4)/4                 \\
V_2 \cos^2 (\theta_0)  &  =  &  (U_1 + U_2 - U_3 - U_4)/4    \label{attr} \\
V_4 \cos^2 (2\theta_0)  &  =  &  (U_1 - U_2 - U_3 + U_4)/4            
\end{eqnarray}
The key point is that, due to the large number of particle-hole excitations 
from the many Fermi points, the interactions with very small 
longitudinal momentum-transfer are more screened than those with some 
transverse (as opposed to longitudinal)
momentum within the same circular patch. This means that 
$U_1 < U_3$, so that an attractive interaction may exist in the Cooper-pair 
channels with $p$-wave ($V_2$) and $d$-wave ($V_4$) symmetry. Moreover, 
as homologous Fermi points in the two circular patches can be connected by 
the fixed vector ${\bf Q}$, the screening is also enhanced for that particular 
momentum. This implies that $U_2 < U_4$, leading from Eq. (\ref{attr}) to 
$V_2 < 0$. This is the origin of the SC instability, as an attractive 
interaction in one of the pairing channels is always driven to strong-coupling 
in the low-energy limit\cite{sh}.

We stress at this point that the disorder introduced by inhomogeneities in the 
electrostatic potential from the different shells or by deformations in a given
nanotube cannot have a significant effect on the SC correlations. A remarkable
property of the graphene lattice is that defects with a spatial size comparable
or larger than the lattice constant give rise to a negligible backscattering in 
the electron propagation\cite{ando}. Then, the effect of such inhomogeneities can 
be modelled by a random potential coupled to the uniform density fluctuations
$\psi^{+}_{a} \psi_{a}$ .
It turns out that the SC correlations are actually insensitive to this kind of
disorder, since the SC order parameter only depends on the variable (momentum)
conjugate to such density fluctuations\cite{gia}. In this perspective, only 
lattice defects with spatial size smaller than the lattice constant may affect
the propagation of the Cooper pairs. Such a kind of disorder may prevent or not 
the development of SC correlations depending on whether it reaches some critical
strength\cite{gia}. This is consistent with the experimental 
observations reported in Ref. \onlinecite{takesue}, where the Raman measurements 
indicate the absence of defects to a large extent in the MWNTs samples.

We have thus developed an approach that is suitable to describe the 
electronic instabilities of doped MWNTs, which may have
many subbands at the Fermi level. This analysis may be relevant to account
for the signatures of SC transitions reported in Ref. \onlinecite{takesue}.
Our approach relies on the scale-dependence of different scattering 
processes at low energies, and it allows to describe the transition from the 
Luttinger liquid to a SC state as some of the Cooper-pair couplings
start to grow large.

We have shown that the mechanism of superconductivity in doped MWNTs is based 
on the particular geometry of the two patches of Fermi points. A reflection of 
this is the appearance of the SC instability in a channel of $p$-wave symmetry. 
It is tempting to relate this feature to the observation of the interplay between 
SC and magnetic order parameters in some samples of doped graphite\cite{moeh}.
All these properties should deserve further investigation at the experimental 
level, for the sake of clarifying as well the relation between the 
superconductivity of different carbon compounds.

The financial support of the Ministerio
de Educaci\'on y Ciencia (Spain) through grant
FIS2005-05478-C02-02 is gratefully acknowledged.
E. P. was also supported by INFN grant 10068.

\end{document}